\RequirePackage{fix-cm}
\documentclass[smallextended]{svjour3}       % onecolumn (second format)
\smartqed  % flush right qed marks, e.g. at end of proof
\usepackage{graphicx}
\usepackage{epstopdf}
\usepackage{amsmath}
\usepackage{bbding}
\begin{document}

\title{A secret sharing scheme on $p^2$-dimensional quantum system%\thanks{Grants or other notes
}


\author{Na Hao         \and
        Zhi-Hui Li         \and
        Hai-Yan Bai         \and
        Chen-Ming Bai %etc.
}


\institute{Na Hao, Zhi-Hui Li\Envelope, Hai-Yan Bai, Chen-Ming Bai \at
            School of Mathematics and Information Science, Shannxi Normal University, Xi'an 710119, China\\
              %Tel.: 13669191125\\
              \email{lizhihui@snnu.edu.cn}           %  \\
           \and
           %Zhi-Hui Li \at
            %  College of Mathematics and Information Science, Shaanxi Normal University, Xi'an, Shaanxi Province CHINA
}

\date{Received: date / Accepted: date}

\maketitle

\begin{abstract}
In this paper, we give the mutually unbiased bases on the $p^2$-dimensional quantum system where $p$ is an odd prime number, and construct the corresponding unitary transformation based on the properties of these mutually unbiased bases. Then, we construct a $(N,N)$ threshold secret sharing scheme using unitary transformation between these mutually unbiased bases, and analyze the scheme's security by several ways, for example, intercept-and-resend attack, entangle-and-measure attack, Trojan horse attack, and so on. Using our method, we construct a single-particle quantum protocol involving only one qudit, and the method shows much more scalability than other schemes.
\end{abstract}

\section{Introduction}
\label{intro}
Secret sharing is an important issue in modern cryptography. It's idea is to divide the secret into several sub-secrets so that the distributor can send the sub-secrets to the participants through the channel, and only participants in authorized subgroup work together can recover the secret, at the same time, non-authorized participants can not get it. The secret sharing schemes can be realized in different ways: the secret sharing schemes based on mathematics is called the classical secret sharing schemes, and the security is ensured  by means of mathematical problems; the secret sharing schemes based on quantum physics is called the quantum secret sharing schemes, and the security is guaranteed with the help of quantum physics law. Hillery et al. [1] proposed the concept of quantum secret sharing with reference to the classical secret sharing schemes, and designed a quantum secret sharing scheme by using the quantum correlation of the GHZ states in 1998. The basic idea of this scheme is to share an unknown quantum state among the two participants. In order to restore the secret, the two parties must cooperate to obtain it. Since then, quantum secret sharing has attracted people's interest, a series of quantum secret sharing schemes have been proposed [2-11].
\par{Mutually unbiased bases (MUBs) is an important tool in many quantum information processing. There have been some results on MUBs [12-18]. In [12], it is proved that the maximum number of MUBs in $d$-dimensional complex space $C^d$ does not exceed $d+1$, and [13] shows that the maximum number of MUBs is $d+1$ when the dimension $d$ of this space is a power of a prime. In [19], Tavakoli et al.  used the cyclicity of MUBs and the theory of unitary matrix to give a secret sharing scheme on $p$-dimensional quantum system where $p$ is an  odd prime. In this paper, we discuss the case of $p^2$-dimensional quantum system with $p$ being an  odd prime. First we study the related properties of MUBs on $C^{p^2}$, and then construct the unitary matrix based on these properties. Finally, we construct a secret sharing scheme on the  $p^2$-dimensional quantum system and discuss the security of the scheme.}
\section{MUBs and their related properties}
\label{sec:1}
\par{We define two bases $B^0=\{\psi^0_1,\psi^0_2,\cdots,\psi^0_d\}$ and $B^1=\{\psi^1_1,\psi^1_2,\cdots,\psi^1_d\}$ over a $d$-dimensional complex space to be mutually  unbiased if the inner products between all possible vector pairs all have the same magnitude:
$$|\langle\psi^0_l|\psi^1_j\rangle|=\frac{1}{\sqrt{d}}\eqno(1)$$
where $l,j=1,2,\cdots,d$. }
\par{{\bf Definition 1} \ \ A set of orthonormal bases $B=\{B^0,B^1,\cdots,B^m\}$ is said to be a set of MUBs if the elements of $B$ are non-biased relative to each other.}
\par{Wootters et al. [13] Pointed out that the maximum number of MUBs in $d$-dimensional complex space is $d+1$ when the dimension $d$ of this space is a power of a prime. We suppose that $d=p^n$, and $p$ is an prime. When $p\neq2$, the specific form of these MUBs is given by using the theory of finite fields. Let $F_{p^n}$ be a finite field, where $p$ is an odd prime, and $n$ is a natural number. We order
$$|v_l^{(j)}\rangle=\frac{1}{\sqrt{d}}\sum_{k\in F_{p^n}}\omega^{{\rm Tr}(jk^2+lk)}|k\rangle,j,l\in F_{p^n},\omega=e^{\frac{2\pi i}{p}}\eqno(2)$$
where $|v_l^{(j)}\rangle\in C^d(d=p^n)$, the superscript $j$ of $|v_l^{(j)}\rangle$ represents the $j$th group base, and the subscript $l$ of $|v_l^{(j)}\rangle$ does the $l$th vector in the base.
When $j\in F_{p^n}$, The bases given by Eq.(2) are denoted in turn $V_0, V_1, \cdots, V_m, \cdots$:}
\par{When $j=0$, $V_0=\{|v_0^{(0)}\rangle,|v_1^{(0)}\rangle,\cdots,|v_l^{(0)}\rangle,\cdots|l\in F_{p^n}\}$;}
\par{When $j=1$, $V_1=\{|v_0^{(1)}\rangle,|v_1^{(1)}\rangle,\cdots,|v_l^{(1)}\rangle,\cdots|l\in F_{p^n}\}$;}
\par{$\cdots$}
\par{When $j=m, m\in F_{p^n}$, $V_m=\{|v_0^{(m)}\rangle,|v_1^{(m)}\rangle,\cdots,|v_l^{(m)}\rangle,\cdots|l\in F_{p^n}\}$;}
\par{$\cdots$}
\par{When $j=d$, we order $$|v_l^{(d)}\rangle=\sum_{k\in F_{p^n}}\delta_{lk}|k\rangle,l\in F_{p^n}.\eqno(3)$$}
\par{We call this group of bases defined by Eq.(3) is a computing base, and the corresponding base is recorded as $V_d=\{|0\rangle,|1\rangle,\cdots,|l\rangle,\cdots|l\in F_{p^n}\}$.}
\par{Let $M=\{V_m|m=0,1,\cdots,d, m\in F_{p^n}\}$. By Definition 1, we can know that $M$ which is consisted of $V_0,V_1,\cdots,V_d$ is a set of MUBs, and the potential of $M$ is $d+1$. When $n=1$, $M$ is a set of  MUBs over $F_p$, which is studied in [16]. Next, we consider the situation of $n=2$. We give the relevant properties of the corresponding  MUBs over $F_{p^2}$, which is defined by the Eq.(2).}
\par{Since $F_{p^2}$ is the quadratic extension of $F_p$, let $f(x)$ be a quadratic irreducible polynomial over $F_p$, and $\theta$ is a root of $f(x)$, then $F_{p^2}=F_p(\theta)=\{k_1+k_2\theta|k_1,k_2\in F_p\}$. Assume $g:F_{p^2}\rightarrow Z_{p^2}$ is a mapping from the finite field $F_{p^2}$ to the residual class ring $Z_{p^2}$, then define
\begin{center}
$g(k_1+k_2\theta)=k_1+k_2\cdot p(mod p^2),k_1+k_2\theta \in F_{p^2}.$
\end{center}
It is easy to verify that $g$ is a one-to-one mapping, and has the following properties.}
\par{{\bf Lemma 1} \ \ Let $F_{p^2}$ be a finite field. The  superscript and the subscript of the MUBs defined by Eq.(2) have the following properties:\\
 1) The subscript $l+1$ of $|v_{l+1}^{(j)}\rangle$ satisfies $l+1=g^{-1}(g(l)+1)({\rm mod} p)$;\\
 2) The superscript $j+1$ of $|v_l^{(j+1)}\rangle$ satisfies $j+1=g^{-1}(g(j)+1)({\rm mod} p)$.}
\par{{\bf Example 1} \ \ when $l=2\theta+p-1$, $j\in F_{p^2}$, where $p$ is a prime number and $p>3$, then we have \begin{align*}
l+1&=g^{-1}(g(l)+1)({\rm mod} p)\\
&=g^{-1}(g(2\theta+p-1)+1)({\rm mod} p)\\
&=g^{-1}((2\cdot p+p-1)+1)({\rm mod} p)\\
&=g^{-1}(3\cdot p)({\rm mod} p)\\
&=3\theta({\rm mod} p)\\
&=3\theta.
\end{align*}}
\begin{flushleft}
Hence, when $l=2\theta+p-1$, $|v_{l+1}^{(j)}\rangle=|v_{2\theta}^{(j)}\rangle, j\in F_{p^2}$.
\end{flushleft}
\par{Further, let $S=\{A_1,A_2,B_1,B_2\}$, where}
\begin{align*}
A_1&=diag(\omega^{{\rm Tr}(0)},\omega^{{\rm Tr}(1)},\cdots,\omega^{{\rm Tr}(p-1)},\omega^{{\rm Tr}(\theta)},\omega^{{\rm Tr}(\theta+1)},\cdots,\omega^{{\rm Tr}(\theta+p-1)},\\
&\cdots,\omega^{{\rm Tr}((p-1)\theta)},\omega^{{\rm Tr}((p-1)\theta+1)},\cdots ,\omega^{{\rm Tr}((p-1)\theta+p-1)});\\
A_2&=diag(\omega^{0a},\omega^{1a},\cdots,\omega^{(p-1)a},\omega^b,\omega^{b+a},\cdots,\omega^{b+(p-1)a},\cdots,\omega^{(p-1)b},\\
&\omega^{(p-1)b+a},\cdots ,\omega^{(p-1)b+(p-1)a});\\
B_1&=diag(\omega^{{\rm Tr}(0^2)},\omega^{{\rm Tr}(1^2)},\cdots,\omega^{{\rm Tr}((p-1)^2)},\omega^{{\rm Tr}(\theta^2)},\omega^{{\rm Tr}((\theta+1)^2)},\cdots,\omega^{{\rm Tr}((\theta+p-1)^2)},\\
&\cdots,\omega^{{\rm Tr}(((p-1)\theta)^2)},\omega^{{\rm Tr}(((p-1)\theta+1)^2)},\cdots ,\omega^{{\rm Tr}(((p-1)\theta+p-1)^2)})\\
B_2&=diag(\omega^{0^2a},\omega^{1^2a},\cdots,\omega^{(p-1)^2a},\omega^{{\rm Tr}(c\theta^2)},\omega^{{\rm Tr}(c(\theta+1)^2)},\cdots,\omega^{{\rm Tr}(c(\theta+p-1)^2)},\\
&\cdots,\omega^{{\rm Tr}(c((p-1)\theta)^2)},\omega^{{\rm Tr}(c((p-1)\theta+1)^2)},\cdots ,\omega^{{\rm Tr}(c((p-1)\theta+p-1)^2)}).
\end{align*}
It is obvious that the order of these four matrixes are all $p$. Here, $\omega=e^{\frac{2\pi i}{p}},a={\rm Tr}(\theta)-{\rm Tr}(p-1),b={\rm Tr}(\theta^2)-{\rm Tr}((p-1)\theta),c=\theta-(p-1)$.
\par{{\bf Theorem 1} \ \ Let $F_{p^2}$ be a finite field, the  MUBs defined by the Eq.(2) have the following properties:\\
 1) When $l\in \{\theta\cdot d+e|d\in Z_p,e\in Z_{p-1}\}, A_1|v_l^{(j)}\rangle=|v_{l+1}^{(j)}\rangle$;\\
 2) When $l\in \{\theta\cdot d+(p-1)|d\in Z_p\}, A_2|v_l^{(j)}\rangle=|v_{l+1}^{(j)}\rangle$;\\
 3) When $j\in \{\theta\cdot d+e|d\in Z_p,e\in Z_{p-1}\}, B_1|v_l^{(j)}\rangle=|v_l^{(j+1)}\rangle$;\\
 4) When $j\in \{\theta\cdot d+(p-1)|d\in Z_p\}, B_2|v_l^{(j)}\rangle=|v_l^{(j+1)}\rangle$.
\par{{\bf Proof} \ \  When $l\in \{\theta\cdot d+e|d\in Z_p,e\in Z_{p-1}\}$,}
\par{$A_1|v_l^{(j)}\rangle$\\
$=A_1\frac{1}{\sqrt{d}}\sum_{k\in F_{p^2}}\omega^{{\rm Tr}(jk^2+lk)}|k\rangle$\\
$=\frac{1}{\sqrt{d}}[\omega^{{\rm Tr}(j\cdot0^2+l\cdot0)}\cdot\omega^{{\rm Tr}(0)}|0\rangle+\omega^{{\rm Tr}(j\cdot1^2+l\cdot1)}\cdot\omega^{{\rm Tr}(1)}|1\rangle+\cdots+\omega^{{\rm Tr}(j\cdot(p-1)^2+l\cdot(p-1))}$\\
$\cdot\omega^{{\rm Tr}(p-1)}|p-1\rangle+\omega^{{\rm Tr}(j\cdot\theta^2+l\cdot\theta)}\cdot\omega^{{\rm Tr}(\theta)}|\theta\rangle+\cdots+\omega^{{\rm Tr}(j\cdot(\theta+p-1)^2+l\cdot(\theta+p-1))}$\\
$\cdot\omega^{{\rm Tr}(\theta+p-1)}|\theta+p-1\rangle+\cdots+\omega^{{\rm Tr}(j\cdot((p-1)\theta)^2+l\cdot((p-1)\theta))}\cdot\omega^{{\rm Tr}((p-1)\theta)}|(p-1)\theta\rangle+\cdots$\\
$+\omega^{{\rm Tr}(j\cdot((p-1)\theta+p-1)^2+l\cdot((p-1)\theta+p-1))}\cdot\omega^{{\rm Tr}((p-1)\theta+p-1)}|(p-1)\theta+p-1\rangle]$\\
$=\frac{1}{\sqrt{d}}[\omega^{{\rm Tr}(j\cdot0^2+(l+1)\cdot0)}|0\rangle+\omega^{{\rm Tr}(j\cdot1^2+(l+1)\cdot1)}|1\rangle+\cdots+\omega^{{\rm Tr}(j\cdot(p-1)^2+(l+1)\cdot(p-1))}$\\
$\cdot|p-1\rangle+\omega^{{\rm Tr}(j\cdot\theta^2+(l+1)\cdot\theta)}|\theta\rangle+\cdots+\omega^{{\rm Tr}(j\cdot(\theta+p-1)^2+(l+1)\cdot(\theta+p-1))}|\theta+p-1\rangle$\\
$+\cdots+\omega^{{\rm Tr}(j\cdot((p-1)\theta)^2+(l+1)\cdot((p-1)\theta))}|(p-1)\theta\rangle+\cdots$\\
$+\omega^{{\rm Tr}(j\cdot((p-1)\theta+p-1)^2+(l+1)\cdot((p-1)\theta+p-1))}|(p-1)\theta+p-1\rangle]$\\
$=\frac{1}{\sqrt{d}}\sum_{k\in F_{p^2}}\omega^{{\rm Tr}(jk^2+(l+1)k)}|k\rangle$\\
$=|v_{l+1}^{(j)}\rangle.$
\par{Therefore, when $l\in \{\theta\cdot d+e|d\in Z_p,e\in Z_{p-1}\}, A_1|v_l^{(j)}\rangle=|v_{l+1}^{(j)}\rangle$. The following 2) 3) and 4) can be proved by the same way:\\
 2) When $l\in \{\theta\cdot d+(p-1)|d\in Z_p\},A_2|v_l^{(j)}\rangle=|v_{l+1}^{(j)}\rangle$;\\
 3) When $j\in \{\theta\cdot d+e|d\in Z_p,e\in Z_{p-1}\},B_1|v_l^{(j)}\rangle=|v_l^{(j+1)}\rangle$;\\
 4) When $j\in \{\theta\cdot d+(p-1)|d\in Z_p\},B_2|v_l^{(j)}\rangle=|v_l^{(j+1)}\rangle$.}
\par{{\bf Remark 1} \ \ The cyclicity of MUBs can be obtained by Theorem 1, as shown in Fig.1 and Fig.2:
\begin{figure*}
  \includegraphics[width=0.75\textwidth]{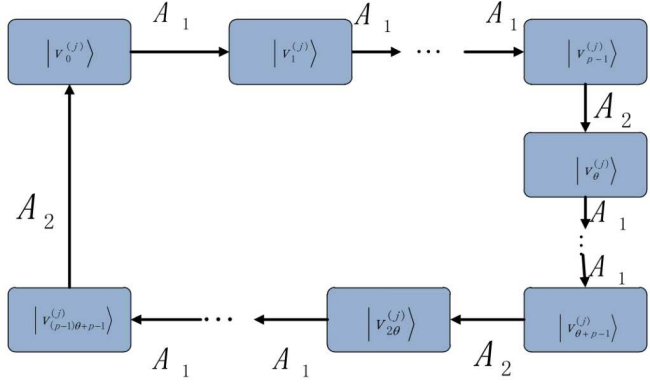}
\caption{cyclicity of the subscript of MUBs.}
\label{fig:1}
\end{figure*}
\begin{figure*}
  \includegraphics[width=0.75\textwidth]{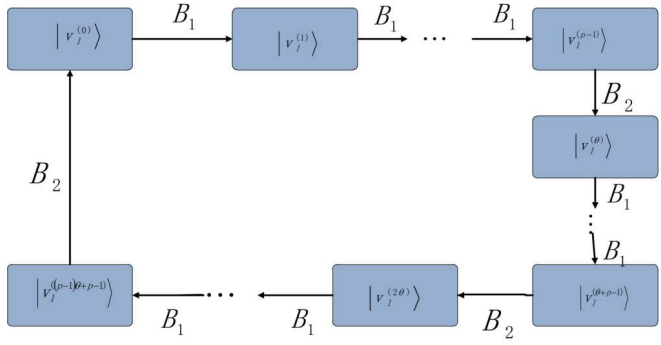}
\caption{cyclicity of the superscript of MUBs.}
\label{fig:2}
\end{figure*}
\section{The construction of the unitary matrix}
\par{In this section, we construct the unitary matrix over $F_{p^2}$ such that for any $\alpha,\beta\in F_{p^2}$, any vector $|v_l^{(j)}\rangle$ can be mapped into $|v_{l+\alpha}^{(j+\beta)}\rangle$. That is, elements of $M$ are mapped into elements of $M$. Note that, any vector $|v_l^{(j)}\rangle$ can be transformed into $|v_{l+\alpha}^{(j+\beta)}\rangle$ by applying the transformation
 $$U_{\alpha,\beta}=A_2^{y_1}A_1^{(p-1)y_1+x_1}B_2^{y_2}B_1^{(p-1)y_2+x_2},\eqno(4)$$
 where $\alpha=x_1+y_1\theta,\beta=x_2+y_2\theta,x_1,x_2,y_1,y_2,\in F_p$, and the addition of $l+\alpha$, $j+\beta$ in $|v_{l+\alpha}^{(j+\beta)}\rangle$ is addition over $Z_p$. The Eq.(4) can be proved as follows:
\begin{flushleft}
Since $A_m,B_m,m=1,2$, are unitary matrixes, and it is easy to be proved that $U_{\alpha,\beta}$ is a unitary matrix. Next we prove that $U_{\alpha,\beta}|v_l^{(j)}\rangle=|v_{l+\alpha}^{(j+\beta)}\rangle$ is held.
\end{flushleft}
}
\par{Assume that $l=k_1+k_2\theta,j=k_3+k_4\theta$, where $k_m\in F_p,m=1,2,3,4$, and $\alpha=x_1+y_1\theta$, $\beta=x_2+y_2\theta$.}
\begin{flushleft}
Hence, $l+\alpha=(k_1+x_1)+(k_2+y_1)\theta, j+\beta=(k_3+x_2)+(k_4+y_2)\theta$.
\end{flushleft}
\begin{flushleft}
By Theorem 1, $|v_l^{(j)}\rangle=A_2^{k_2}A_1^{(p-1)k_2+k_1}B_2^{k_4}B_1^{(p-1)k_4+k_3}|v_0^{(0)}\rangle$, then we have
\end{flushleft}
\par{$U_{\alpha,\beta}|v_l^{(j)}\rangle$\\
$=A_2^{y_1}A_1^{(p-1)y_1+x_1}B_2^{y_2}B_1^{(p-1)y_2+x_2}\cdot(A_2^{k_2}A_1^{(p-1)k_2+k_1}B_2^{k_4}B_1^{(p-1)k_4+k_3}|v_0^{(0)}\rangle)$\\
$=A_2^{y_1}A_1^{(p-1)y_1+x_1}B_2^{y_2}B_1^{(p-1)y_2+x_2}A_2^{k_2}A_1^{(p-1)k_2+k_1}B_2^{k_4}B_1^{(p-1)k_4+k_3}|v_0^{(0)}\rangle$\\
$=A_2^{y_1+k_2}A_1^{(p-1)(y_1+k_2)+(x_1+k_1)}B_2^{y_2+k_4}B_1^{(p-1)(y_2+k_4)+(x_2+k_3)}|v_0^{(0)}\rangle$\\
$=|v_{l+\alpha}^{(j+\beta)}\rangle$.}
\begin{flushleft}
Hence, $U_{\alpha,\beta}|v_l^{(j)}\rangle=|v_{l+\alpha}^{(j+\beta)}\rangle$.
\end{flushleft}
\section{$(N,N)$threshold scheme}
\par{In this section, we construct a $(N,N)$ threshold secret sharing scheme over $F_{p^2}$. Here ${\rm Alice}$ is a distributor, and ${\rm Bob_1,\rm Bob_2,\cdots,\rm Bob_N}$ are the participants.}
\subsection{Preparation phase}
\par{\ \ \ \  (1)${\rm Alice}$ prepares quantum state $|v_0^{(0)}\rangle$, where $|v_0^{(0)}\rangle=\frac{1}{\sqrt{p^2}}\sum_{k\in F_{p^2}}|k\rangle$;}
\par{(2)${\rm Alice}$ randomly selects independent $\alpha_0,\beta_0\in F_{p^2}$, and lets $\alpha_0=x_{01}+y_{01}\theta, \beta_0=x_{02}+y_{02}\theta$, where $x_{01}, y_{01}, x_{02}, y_{02}\in F_p$. By Eq.(4), $U_{\alpha_0,\beta_0}=A_2^{y_{01}}A_1^{(p-1)y_{01}+x_{01}}B_2^{y_{02}}B_1^{(p-1)y_{02}+x_{02}}$, ${\rm Alice}$ performs on $|v_0^{(0)}\rangle$ the transformation $U_{\alpha_0,\beta_0}$. Her action gives a state $|\psi_0\rangle$ which is sent to ${\rm Bob_1}$;}
\par{(3)${\rm Alice}$ randomly selects independent $\beta_1, \beta_2, \cdots, \beta_N\in F_{p^2}$, and sends $\beta_k$ to ${\rm Bob}_k$, where $\beta_k=x_{k2}+y_{k2}\theta$, $x_{k2}, y_{k2}\in F_p$ for $k=1, 2, \cdots, N$.}
\subsection{Distribution phase}
\par{\ \ \ \  (1)${\rm Bob}_1$ randomly chooses $\alpha_1\in F_{p^2}$, and lets $\alpha_1=x_{11}+y_{11}\theta$, where $x_{11},y_{11}\in F_p$, and $\beta_1=x_{12}+y_{12}\theta$ received from ${\rm Alice}$. By Eq.(4), $U_{\alpha_1,\beta_1}=A_2^{y_{11}}A_1^{(p-1)y_{11}+x_{11}}B_2^{y_{12}}B_1^{(p-1)y_{12}+x_{12}}$, ${\rm Bob}_1$ applies $U_{\alpha_1,\beta_1}$ to the qudit $|\psi_0\rangle$ received from ${\rm Alice}$. ${\rm Bob}_1$'s action gives a state $|\psi_1\rangle$, which is sent ${\rm Bob}_2$;}
\par{(2)${\rm Bob}_2$ randomly chooses $\alpha_2\in F_{p^2}$, and lets $\alpha_2=x_{21}+y_{21}\theta$, where $x_{21},y_{21}\in F_p$, and $\beta_2=x_{22}+y_{22}\theta$ received from ${\rm Alice}$. By Eq.(4), $U_{\alpha_2,\beta_2}=A_2^{y_{21}}A_1^{(p-1)y_{21}+x_{21}}B_2^{y_{22}}B_1^{(p-1)y_{22}+x_{22}}$, ${\rm Bob}_2$ applies $U_{\alpha_2,\beta_2}$ to the qudit $|\psi_1\rangle$ received from ${\rm Bob}_1$. ${\rm Bob}_2$'s action gives a state $|\psi_2\rangle$, which is sent ${\rm Bob}_3$;}
\par{(3)Repeat this process until ${\rm Bob}_N$ receives the state $|\psi_{N-1}\rangle$ from ${\rm Bob}_{N-1}$. ${\rm Bob}_N$ randomly selects $\alpha_N\in F_{p^2}$, and lets $\alpha_N=x_{N1}+y_{N1}\theta$, where $x_{N1}$, $y_{N1}\in F_p$, and $\beta_N=x_{N2}+y_{N2}\theta$ received from ${\rm Alice}$. By Eq.(4),
 $U_{\alpha_N,\beta_N}=A_2^{y_{N1}}A_1^{(p-1)y_{N1}+x_{N1}}B_2^{y_{N2}}B_1^{(p-1)y_{N2}+x_{N2}}$, ${\rm Bob}_N$ applies $U_{\alpha_N,\beta_N}$ to the qudit $|\psi_{N-1}\rangle$. ${\rm Bob}_N$'s action gives a state $|\psi_N\rangle$, which is sent ${\rm Alice}$.}
\subsection{Measurement phase}
\par{${\rm Alice}$ calculates the $\sum_{k=0}^N \beta_k({\rm mod} p)$ privately, and records the results as $J$, and uses the base $|v_l^{(J)}\rangle$ to measure the quantum state $|\psi_N\rangle$. If the result is $|v_h^{(J)}\rangle$, then ${\rm Alice}$ marks $h,h\in F_{p^2}$.}
\subsection{Testing phase}
\par{In order to check the security, ${\rm Alice}$ orders ${\rm Bob}_k$ announce their choice of $\alpha_k$ for $k=1,2,\cdots,N$, and  calculates whether the following formula holds:
$$\sum_{k=0}^N \alpha_k=h({\rm mod} p).\eqno(5)$$
If the Eq.(5) holds, this run is safe, and the protocol continues; otherwise, it is unsafe, and there is eavesdropping or deceit. Then we discard this run.}
\subsection{Recovery phase}
\par{ If this run is safe, parties ${\rm Bob}_1,{\rm Bob}_2,\cdots,{\rm Bob}_N$, after changing all their data $\alpha_k$ for a valid run, not used in the security check, can reconstruct the key $s$ for the given run:
$$s=\sum_{k=1}^N \alpha_k({\rm mod} p).$$}
\par{{\bf Remark 2} The reason why the protocol can run is that:\begin{center}
$|\psi_N\rangle=U_{\alpha_N,\beta_N}\cdot U_{\alpha_{N-1},\beta_{N-1}}\cdot \cdots$
$\cdot U_{\alpha_0,\beta_0}|v_0^{(0)}\rangle$
$=|v_{\Delta}^{(\Gamma)}\rangle.$
\end{center}
where $\Delta=\sum_{k=0}^N \alpha_k({\rm mod} p^2)$, $\Gamma=\sum_{k=0}^N \beta_k({\rm mod} p^2)$.}
\section{Security analysis}
\par{The secret sharing scheme must ensure security. In this section, we discuss the security of the scheme in the following two aspects.}
\subsection{External attack}
\subsubsection{Intercept-and-resend attack}
\par{If there is an external eavesdropper Eve, she tries to take an intercept-and-resend attack and intercept a quantum state $|v_l^{(j)}\rangle$ sent from ${\rm Bob}_k$ to ${\rm Bob}_{k+1}$. She selects one of $d$ relevant bases to measure, where $d=p^2$. Eve attacks successfully with a probability of $\frac{1}{d}$, However, with a probability of $\frac{d-1}{d}$ she fails. When the attack fails, the quantum state that Eve resends to ${\rm Bob}_{k+1}$ will be changed and the eavesdropping will be found. Therefore, the success of eavesdropping depends largely on $d$, and inconsistency will be caused between the private data and Eq.(5), so the protocol testing state can be checked out.}
\subsubsection{More general eavesdropping attacks}
\par{For the more general eavesdropping attacks, in the qudit transfer from ${\rm Bob}_k$ to ${\rm Bob}_{k+1}$, we can treat the parties ${\rm Bob}_1,\cdots,{\rm Bob}_k$ as a "block" effectively representing a single community, and parties ${\rm Bob}_{k+1},\cdots,{\rm Bob}_{N+1}$ and ${\rm Bob}_1$, acting as a measurement community, so the eavesdropping attack is simplified to the case of previously encountered in the BB84 two-party QKD, we can treat it in a similar way.}
\subsubsection{Entangle-and-measure attack}
\par{If there is an external eavesdropper Eve, she tries to take an entangle-and-measure attack and uses a unitary operation $U_E$ to entangle an ancillary particle on the transmitted particle, then she measures the ancillary particle to steal secret information. Assume that the ancillary particle is $|E\rangle$. Eve performs the unitary transform $U_E$ on her particles and $|E\rangle$ in the following forms, $$U_E|k\rangle|E\rangle=\sum_{l\in F_{p^2}}a_{kl}|l\rangle|e_{kl}\rangle.\eqno(6)$$
\begin{align*}
U_E|v_l^{(j)}\rangle|E\rangle&=U_E(\frac{1}{\sqrt{p^2}}\sum_{k\in F_{p^2}}\omega^{{\rm Tr}(jk^2+lk)}|k\rangle)|E\rangle\\
&=\frac{1}{\sqrt{p^2}}\sum_{k\in F_{p^2}}\omega^{{\rm Tr}(jk^2+lk)}U_E|k\rangle|E\rangle\\
&=\frac{1}{\sqrt{p^2}}\sum_{k\in F_{p^2}}\omega^{{\rm Tr}(jk^2+lk)}(\sum_{l\in F_{p^2}}a_{kl}|l\rangle|e_{kl}\rangle)\\
&=\frac{1}{\sqrt{p^2}}\sum_{k\in F_{p^2}}\sum_{l\in F_{p^2}}\omega^{{\rm Tr}(jk^2+lk)}a_{kl}(\frac{1}{\sqrt{p^2}}\sum_{m\in F_{p^2}}\omega^{-{\rm Tr}(jl^2+ml)}|v_m^{(j)}\rangle)|e_{kl}\rangle\\
&=\frac{1}{p^2}\sum_{k\in F_{p^2}}\sum_{l\in F_{p^2}}\sum_{m\in F_{p^2}}\omega^{{\rm Tr}(jk^2+lk)-{\rm Tr}(jl^2+ml)}a_{kl}|v_m^{(j)}\rangle|e_{kl}\rangle.
\end{align*}
that is to say, $$U_E|v_l^{(j)}\rangle|E\rangle=\frac{1}{p^2}\sum_{k\in F_{p^2}}\sum_{l\in F_{p^2}}\sum_{m\in F_{p^2}}\omega^{{\rm Tr}(jk^2+lk)-{\rm Tr}(jl^2+ml)}a_{kl}|v_m^{(j)}\rangle|e_{kl}\rangle.\eqno(7)$$
where $|v_l^{(j)}\rangle=\frac{1}{\sqrt{p^2}}\sum_{k\in F_{p^2}}\omega^{{\rm Tr}(jk^2+lk)}|k\rangle, \omega=e^{\frac{2\pi i}{p}}, k,l,j,m\in F_{p^2}$, $|e_{kl}\rangle$ is the pure auxiliary state determined uniquely by the unitary transform $U_E$, and $$\sum_{l\in F_{p^2}}|a_{kl}|^2=1(k\in F_{p^2}).\eqno(8)$$}
\par{In order to avoid wiretapping detection, Eve has to set: $a_{kl}=0$, where $k\neq l$ and $k,l\in F_{p^2}$. Therefore, the Eq.(6) and Eq.(7) can be simplified as follows:
$$U_E|k\rangle|E\rangle=a_k|k\rangle|e_k\rangle.\eqno(9)$$
$$U_E|v_l^{(j)}\rangle|E\rangle=\frac{1}{p^2}\sum_{k\in F_{p^2}}\sum_{m\in
F_{p^2}}\omega^{{\rm Tr}((l-m)k)}a_k|v_m^{(j)}\rangle|e_k\rangle.\eqno(10)$$
where $a_k=a_{kk},e_k=e_{kk}$.}
\par{Similarly, Eve can obtain that $\sum_{k\in F_{p^2}}\omega^{{\rm Tr}((l-m)k)}a_k|e_k\rangle=0$, where $m\neq l$, and $m\in F_{p^2}$. Then for any $l\in F_{p^2}$, we can get $p^2-1$ equations. According to these $p^2-1$ equations, we can compute that $$a_0|e_0\rangle=a_1|e_1\rangle=\cdots=a_m|e_m\rangle=\cdots=a_{(p-1)\theta+p-1}|e_{(p-1)\theta+p-1}\rangle,m\in F_{p^2}.$$ }
\par{To obtain useful information about the secret, without loss of generality, Eve uses the unitary transform $U_E$ on the $|v_0^{(0)}\rangle=\frac{1}{\sqrt{p^2}}(|0\rangle+|1\rangle+\cdots+|(p-1)\theta+p-1\rangle)$, that is to say,}
\par{$U_E|v_0^{(0)}\rangle$
\begin{flushleft}
$=\frac{1}{\sqrt{p^2}}(a_0|0\rangle|e_0\rangle+a_1|1\rangle|e_1\rangle+\cdots+a_{(p-1)\theta+p-1}|(p-1)\theta+p-1\rangle|e_{(p-1)\theta+p-1}\rangle)$
$=\frac{1}{\sqrt{p^2}}(|0\rangle+|1\rangle+\cdots+|(p-1)\theta+p-1\rangle)\otimes(a_0|e_0\rangle)$.
\end{flushleft}
}
\par{It is easy to see from the above equation that if Eve wants to avoid eavesdropping detection, she can not affect the whole system of quantum secret sharing, and can not get effective information from the auxiliary particle. Therefore, the entanglement measurement attack is invalid.}
\subsubsection{Trojan horse attack}
\par{In this section, we will primarily consider another important attack-Trojan horse attack. If the particles used in the QSS scheme are photons, the proposed protocol may be unsafe for the following two Trojan horse attacks: the delay photon attack and the invisible photon attack. In order to prevent the delay photon attack, the participants must be able to distinguish whether there are multi-photon signals. That is to say, it must be able to distinguish the received photons from single-photon and multi-photon. PNS technology can meet this requirement. The participants can pick up a portion of the photons and split each particle by the PNS technology, then they measure the photons. If the multi-photon rate is much higher than the desired value, the transfer process should be stopped and restarted. To prevent the invisible photon attacks, the participants need to add filters to the device. The filter only allows photons whose wavelength of the photon signal is close to the wavelength of the operating particle to enter. Therefore, the attacker's invisible photons will be isolated.}
\subsection{Internal attack}
\par{In a quantum secret sharing scheme, it is very important to defend against the participant's conspiracy attack because it is more simple and effective than other external attacks. In fact, participant conspiratorial attacks have destroyed many quantum secret sharing schemes. Here, we will discuss the security of our scheme against some particular conspiracies. However, rigorous security proof for general conspiracies is unknown.}
\subsubsection{Only one dishonest participant}
\par{In this section, we suppose that there is only one dishonest participant ${\rm Bob}_j$ for $j\in{1,2,\cdots,N}$ who receives $|\psi_{j-1}\rangle$ from ${\rm Bob}_{j-1}$ when $j>1$ or ${\rm Alice}$ when $j=1$, then ${\rm Bob}_j$ applies $U_{\alpha_j,\beta_j}$ to the qudit $|\psi_{j-1}\rangle$, and passes it to the next participant. In the testing phase, when ${\rm Bob}_j$ publishes the data $\alpha_j$, two operations can be performed: (a) ${\rm Bob}_j$ announces the wrong $\alpha_j$, which can be directly judged by Eq.(5) that this run of protocol is unsafe and we discard this run; (b) ${\rm Bob}_j$ announces the correct $\alpha_j$. In this case, Eq.(5) holds, and then this run will not be detected the exist of the dishonest participant, and the protocol continues to be implemented. Then, dishonest ${\rm Bob}_j$ can correctly guess the value of $\sum_{n=1}^N\alpha_n-\alpha_j$ with probability $\frac{1}{d}$, where $d=p^2$, and then get the key $\sum_{n=1}^N\alpha_n$. But with probability $\frac{d-1}{d}$ he guesses wrong in which case the key can not be obtained accurately, and the eavesdropping fails. When applying the scheme, we always carry out multiple rounds for $n$ rounds of the above agreement, not only one round, and then the probability of successful eavesdropping $(\frac{1}{d})^n$ will tend to zero, that is to say, the protocol is safe.}
\subsubsection{$k$ dishonest participants}
\par{Assuming that there are $k$ dishonest participants $\alpha_{j_1},\alpha_{j_2},\cdots,\alpha_{j_k}$. Similar to only one dishonest participant, consider the $k$ individuals as a whole $K$ where $\alpha_K=\alpha_{j_1}+\alpha_{j_2}+\cdots+\alpha_{j_k}$. In the testing phase, when $K$ publishes the data $\alpha_K$, two operations can be performed: (a) $K$ announces the wrong $\alpha_K$, which can be directly judged by Eq.(5) that this run of protocol is unsafe and we discard this run; (b) $K$ announces the correct $\alpha_K$. In this case, Eq.(5) holds, and then this run will not be detected the exist of the dishonest participant, and the protocol continues to be implemented. Then, dishonest $K$ can correctly guess the value of $\sum_{n=1}^N\alpha_n-\alpha_K$ with probability $\frac{1}{d}$, where $d=p^2$, and then get the key $\sum_{n=1}^N\alpha_n$. But with probability $\frac{d-1}{d}$ he guesses wrong in which case the key can not be obtained accurately, and the eavesdropping fails. When applying the scheme, we always carry out multiple rounds for $n$ rounds of the above agreement, not only one round, and then the probability of successful eavesdropping $(\frac{1}{d})^n$ will tend to zero, that is, the protocol is safe.}
\subsubsection{Using quantum memories and entangling of systems with an ancilla}
\par{In,e.g.,Refs.[20] eavesdropping attacks use quantum memories and entangling of systems with an ancilla. However, the attacks of [20] require that either the first or the final party are cheating, which never happens in our protocol because ${\rm Alice}$ is effectively both first and last party. Additionally, the eavesdropping attacks of [20] require knowledge of also $\beta_0$ and $J$, which is impossible since ${\rm Alice}$ never announces any data.}
\par{To facilitate the understanding of the above scheme, we give an example as follows:}
\par{{\bf Example 2} We consider the $(3,3)$ threshold secret sharing scheme for $p=3$, $d=3^2$. That is, there are one distributor ${\rm Alice}$ and three participants ${\rm Bob}_1$, ${\rm Bob}_2, {\rm Bob}_3$.}
\par{When $p=3, d=3^2$, we can get $U_{\alpha,\beta}=A_2^{y_1}A_1^{2y_1+x_1}B_2^{y_2}B_1^{2y_2+x_2}$ by Eq.(4) where $\alpha, \beta\in F_9$. We suppose that $f(x)=x^2+x+2$, then we can calculate that $a={\rm Tr}(\theta)-{\rm Tr}(p-1)=1$, $b={\rm Tr}(\theta^2)-{\rm Tr}((p-1)\theta)=2$, $c=\theta-(p-1)=\theta-2$, and
\begin{align*}
A_1&=diag(1,\omega^2,\omega,\omega^2,\omega,1,\omega,1,\omega^2);\\
A_2&=diag(1,\omega,\omega^2,\omega^2,1,\omega,\omega,\omega^2,1);\\
B_1&=diag(1,\omega^2,\omega^2,1,1,\omega,1,\omega,1);\\
B_2&=diag(1,\omega,\omega,\omega^2,\omega,\omega^2,\omega^2,\omega^2,\omega).
\end{align*}}
\par{{\bf Preparation phase:}}

(1)${\rm Alice}$ prepares the quantum state $|v_0^{(0)}\rangle$:
\begin{align*}
|v_0^{(0)}\rangle&=\frac{1}{\sqrt{p^2}}\sum_{k\in F_{p^2}}|k\rangle\\
&=\frac{1}{3}(1,1,1,1,1,1,1,1,1)^\prime;
\end{align*}
\par{(2)${\rm Alice}$ randomly selects independent $\alpha_0,\beta_0\in F_9$. We assume that $\alpha_0=1+\theta,\beta_0=\theta$. By Eq.(4), we can know that $x_{01}=1,y_{01}=1,x_{02}=0,y_{02}=1$, and \begin{align*} U_{\alpha_0,\beta_0}&=A_2^{y_{01}}A_1^{2y_{01}+x_{01}}B_2^{y_{02}}B_1^{2y_{02}+x_{02}}\\
&=A_2^1A_1^{2\cdot1+1}B_2^1B_1^{2\cdot1+0}\\
&=A_2B_2B_1^2.
\end{align*}
$|\psi_0\rangle=U_{\alpha_0,\beta_0}|v_0^{(0)}\rangle=A_2B_2B_1^2|v_0^{(0)}\rangle=\frac{1}{3}(1,1,\omega,\omega,\omega,\omega^2,1,1,\omega)^\prime$, and $|\psi_0\rangle$ is passed to ${\rm Bob_1}$;}
\par{(3)${\rm Alice}$ randomly selects independent $\beta_1,\beta_2,\beta_3\in F_9$, and sends $\beta_k$ to ${\rm Bob}_k$ for $k=1,2,3$. We assume that $\beta_1=1,\beta_2=\theta,\beta_3=1+\theta$.}

   {\bf Distribution phase:}

(1)${\rm Bob}_1$: ${\rm Bob}_1$ randomly selects $\alpha_1\in F_9$, and we assume that $\alpha_1=\theta$. At the same time, $\beta_1=1$. By Eq.(4), we can know that $x_{11}=0, y_{11}=1, x_{12}=1, y_{12}=0$ and \begin{align*} U_{\alpha_1,\beta_1}&=A_2^{y_{11}}A_1^{(p-1)y_{11}+x_{11}}B_2^{y_{12}}B_1^{(p-1)y_{12}+x_{12}}\\
&=A_2^1A_1^{2\cdot1+0}\cdot B_2^0B_1^{2\cdot0+1}\\
&=A_2A_1^2B_1.
\end{align*}
$|\psi_1\rangle=U_{\alpha_1,\beta_1}|\psi_0\rangle=A_2A_1^2B_1|\psi_0\rangle=\frac{1}{3}(1,\omega,\omega,\omega,1,\omega,1,1,\omega^2)^\prime$,
 and $|\psi_1\rangle$ is passed to ${\rm Bob}_2$;
\par{(2)${\rm Bob}_2$: ${\rm Bob}_2$ randomly selects $\alpha_2\in F_9$, and we assume that $\alpha_2=\theta$. At the same time, $\beta_2=\theta$. By Eq.(4), we can know that $x_{21}=0, y_{21}=1, x_{22}=0, y_{22}=1$ and \begin{align*} U_{\alpha_2,\beta_2}&=A_2^{y_{21}}A_1^{(p-1)y_{21}+x_{21}}B_2^{y_{22}}B_1^{(p-1)y_{22}+x_{22}}\\
&=A_2^1A_1^{2\cdot1+0}\cdot B_2^1B_1^{2\cdot1+0}\\
&=A_2A_1^2B_2B_2^2.
\end{align*}
$|\psi_2\rangle=U_{\alpha_2,\beta_2}|\psi_1\rangle=A_2A_1^2B_2B_1^2|\psi_1\rangle=\frac{1}{3}(1,\omega^2,\omega,1,1,1,\omega^2,1,\omega)^\prime$,
 and $|\psi_2\rangle$ is passed to ${\rm Bob}_3$;}
\par{(3)${\rm Bob}_3$: ${\rm Bob}_3$ randomly selects $\alpha_3\in F_9$, and we assume that $\alpha_3=2+\theta$. At the same time, $\beta_3=1+\theta$. By Eq.(4), we can know that $x_{31}=2, y_{31}=1, x_{32}=1, y_{32}=1$ and \begin{align*} U_{\alpha_3,\beta_3}&=A_2^{y_{31}}A_1^{(p-1)y_{31}+x_{31}}B_2^{y_{32}}B_1^{(p-1)y_{32}+x_{32}}\\
&=A_2^1A_1^{2\cdot1+2}\cdot B_2^1B_1^{2\cdot1+1}\\
&=A_2A_1B_2.
\end{align*}
$|\psi_3\rangle=U_{\alpha_3,\beta_3}|\psi_2\rangle=A_2A_1B_2|\psi_2\rangle=\frac{1}{3}(1,1,\omega^2,1,\omega^2,1,1,\omega,\omega)^\prime$,
 and $|\psi_3\rangle$ is passed to ${\rm Alice}$.}

 {\bf Measurement phase:}

${\rm Alice}$ calculates the $\sum_{k=0}^3 \beta_k({\rm mod} 3)=\theta+1+\theta+(1+\theta)=2$ privately, and uses the base $|v_l^{(2)}\rangle$ to measure the quantum state $|\psi_3\rangle$. The result is $|v_\theta^{(2)}\rangle$, then ${\rm Alice}$ marks $\theta,\theta\in F_9$.

 {\bf Testing phase:}

In order to check the security, ${\rm Alice}$ lets the participants ${\rm Bob}_k$ announce their choice of $\alpha_k$ for $k=1,2,3$, and calculates  the following formula:
$$\sum_{k=0}^3 \alpha_k=(1+\theta)+\theta+\theta+(2+\theta)=\theta({\rm mod} 3)=\theta.$$
Eq.(5) holds, this run is safe, and the protocol continues.

 {\bf Recovery phase:}

${\rm Bob}_k$ share their $\alpha_k$ for $k=1,2,3$, and they will get the key: $$s=\sum_{k=1}^3 \alpha_k({\rm mod} 3)=\theta+\theta+(2+\theta)({\rm mod} 3)=2.$$
So, the key generated by this round is $2$.
\section{Conclusions}
\par{Using our method, we construct a single-particle quantum protocol involving only one qudit. Although our protocol does not consider possible complex attacks, it can resist standard attacks and show much more scalability than other schemes. Compared to the [19], we extend the case where $d$ is an odd prime to the square of an odd prime. Of course, we guess that the result can be generalized to any power of odd prime, but these cases are more complex. This is the question that we will further study in the future.}
%\begin{acknowledgements}
%\end{acknowledgements}


\begin{thebibliography}{00}
\bibitem{Hillery}
Hillery, M., Buzek, V., Berthiaume, A., Quantum secret sharing, Phys. Rev. A 59, 1829 (1999)
\bibitem{Bai2017}
Bai, C. M., Li, Z. H., Liu, C. J., et al, Quantum secret sharing using orthogonal multiqudit entangled states, Quantum Inf Process, 16, 304 (2017)
\bibitem{Wang}
Wang, J. T., Li, L. X., Peng, H. P., et al., Quantum-secret-sharing scheme based on local distinguishability of orthogonal multiqudit entangled states, Phys. Rev. A  95, 022320 (2017)
\bibitem{Bai17}
Bai, C. M., Li, Z. H., Xu, T. T., et al, Quantum secret sharing using the d-dimensional GHZ state, Quantum Inf Process, 16, 59 (2017)
\bibitem{Qin}
Qin, H. W., Dai, Y. W., Verifiable $(t,n)$ threshold quantum secret sharing using $d$-dimensional Bell state, Information Processing Letters 116, 351-355 (2016)
\bibitem{Bai16}
Bai, C. M., Li, Z. H., Xu, T. T., Li, Y. M., A Generalized Information Theoretical Model for Quantum Secret Sharing, Int J Theor Phys, 55, 4972-4986 (2016)
\bibitem{Rahaman}
Rahaman, R., Parker, M. G., Quantum scheme for secret sharing based on local distinguishability, Phys. Rev. A 91, 022330 (2015)
\bibitem{Huang}
Huang, D. Z., Chen, Z. G., Guo, Y., Multiparty Quantum Secret Sharing Using Quantum Fourier Transform, Commun. Theor. Phys. (Beijing, China) 51, 221-226 (2009)
\bibitem{Yu}
Yu, I.C., Lin, F. L., Huang, C. Y., Quantum secret sharing with multilevel mutually (un)biased bases, Phys. Rev. A 78, 012344 (2008)
\bibitem{Singh}
Singh, S. K., Srikanth, R., Generalized quantum secret sharing, Phys. Rev. A  71, 012328 (2005)
\bibitem{Nascimento}
Anderson, C. A. N., Joern, M. Q., Hideki, I., Improving quantum secret-sharing schemes, Phys. Rev. A 64, 042311 (2001)
\bibitem{Calderbank}
Calderbank, A. R., Cameron, P. J., Kantor, W. M., et al., $Z_4$-Kerdock Codes, Orthogonal sperads, and extremal euclidean line-sets, Proc. London Math. Soc. 75, 436-480 (1997)
\bibitem{Wootters}
Wootters, W. K., Fields, B. D., Optimal state-determination by mutually unbiased measurements, Ann. Phys. (N.Y) 191, 363-381 (1989)
\bibitem{Chen}
Chen, L., Yu, L., Product states and Schmidt rank of mutually unbiased bases in dimension six, J. Phys. A: Math. Theor. 50, 475304 (31pp) (2017)
\bibitem{Ghiu}
Ghiu, I., Generation of all sets of mutually unbiased bases for three-qubit systems, Phys. Scr. T153, 014027 (5pp) (2013)
\bibitem{Wiesniak}
Wiesniak, M, Paterek, T, Zeilinger, A, Entanglement in mutually unbiased bases,  New J. Phys. 13, 053047 (2011)
\bibitem{Klimov}
Klimov, A. B, Romero, J. L, Bjork, G., et al., Geometrical approach to mutually unbiased bases, J. Phys. A: Math. Theor. 40, 3987¨C3998 (2007)
\bibitem{Lawrence}
Lawrence, J., Brukner, C., Zeilinger, A., Mutually unbiased binary observable sets on $N$ qubits, Phys. Rev. A 65, 032320 (2002)
\bibitem{Tavakoli}
Tavakoli, A., Herbauts, I., Zukowski, M., et al., Secret sharing with a single d-level quantum system, Phys. Rev. A  92, 030302(R) (2015)
\bibitem{He}
He, G. P., Wang, Z. D., Single qubit quantum secret sharing with improved security, Quantum Inf. Comput. 10, 28 (2010)
\end{thebibliography}


\end{document}